\documentclass[showpacs,aps,preprint,prc]{revtex4}
\usepackage{graphicx}

\begin{document}
\title{\bf The singular inverse square potential, limit cycles and self-adjoint
extensions}

\author{M.\ Bawin}

\affiliation{Universit\'{e} de Li\`{e}ge, Institut de Physique B5, Sart
Tilman, 4000 Li\`{e}ge 1, Belgium  }

\author{S.\ A.\ Coon}
\affiliation{Physics Department, New Mexico State University, Las Cruces, NM 
88003, USA and National Science Foundation, 4201 Wilson Blvd., Arlington,
Virginia 22230, USA}

\date{\today}

\begin{abstract}

We study the radial Schroedinger equation for a particle  of mass m in the
field of a singular attractive $ {\alpha}/{r^2}$ potential with $2m\alpha >
{1}/{4}$. This potential is relevant to the fabrication of nanoscale atom optical devices, is said to be the potential describing the dipole-bound anions of polar molecules, and is the effective potential underlying the universal behavior of three-body systems in nuclear physics and atomic physics, including aspects of Bose-Einstein condensates, first described by Efimov. New results in  three-body physical systems motivate the present investigation. Using the regularization method of Beane et al., we show that the
corresponding ``renormalization group flow'' equation can be solved
analytically. We find that it exhibits a limit cycle behavior and has
infinitely many branches. We show that a physical meaning for self-adjoint 
extensions of the Hamiltonian arises naturally in this framework.
\end{abstract}

\pacs{ 34.20.-b, 03.65.Ca, 03.65.Ge, 11.10.Hi,} 

%\vskip20pt

\maketitle

In this note, we study the regularization and renormalization of the  singular
attractive $ {\alpha}/{r^2}$ potential, a problem motivated, in part, by
recent results in three-body physics.  Specifically, there has been renewed
interest in the non-relativistic three-body system with short-range
interactions. The current investigations are stimulated by the promise that
effective field theories (EFT'S) offer a systematic and model-independent
treatment of atomic, nuclear and hadronic physics at low energies. That is, a
low-energy system with a clear-cut separation of distance scales can be
described by an EFT involving explicitly only the long-wavelength degrees of
freedom. The short-range dynamics can be treated as a set of local operators
which correspond to delta-function  interactions in coordinate space. The
details of the short distance physics cannot be of importance to the low energy
aspects of the system; if they are, it is an indication of a need for
renormalization of the EFT. 

Such a renormalization of an EFT of a three-body system with delta-function 
two-body interactions \cite{BHK} has lead to the rediscovery of the
one-parameter contact three-body interaction shown to restore the lower bound
of this three-body Hamiltonian \cite{AHW}. (Here we mention that the
unboundedness of the Hamiltonian from below is interpreted in EFT's as the
onset of short-distance physics whose effect must be included in local
counterterms \cite{Beane}; hence the three-body counterterm in the EFT
equations for 3-body systems, both bound and scattering states \cite{EFTmisc}.)
The three-body counterterm exhibits a convergence of the renormalization group
flow to one-dimensional limit cycles. This was earlier proved in Ref.
\cite{AHW}, a mathematical analysis of the Efimov effect which occurs in bound
3-body systems when more than one of the two-body subsystems has a zero energy
resonance \cite{Efi}. The presumed novelty of a renormalization group flow with
a limit cycle has inspired additional recent work \cite{Wil}.
 
As the applications of EFT's continue \cite{EFTmisc,Bra}, it will be important
to understand how to explicitly renormalize higher orders in an EFT. For
example, a renormalized equation for two-nucleon systems with explicit
pion-exchange would be of great potential value. Pion exchange gives rise to a
singular $ 1/r^3$ potential and the questions arise: Can the resummation of
pion graphs be renormalized by a single local operator? Would this operator
exhibit a limit cycle as does the three-body contact operator in the pion-less
three-nucleon EFT? Could one calculate the evolution of such an operator
analytically? A positive answer to the last question would help future
numerical work with EFT's. Already, the short distance physics of the $^3S_1$ coupled channels of the single pion exchange potential has been renormalized by a short range four-nucleon counterterm using the method introduced in ref.[3], but the treatment was numerical and these questions were not addressed in that investigation \cite{BeBe}

Such questions, coupled with the interesting limit cycle behavior found in the
three-body system, have prompted investigations of the renormalization group
behavior of the short range counterterms which serve to regularize given long
range potentials (including singular potentials) in the two-body Schroedinger
equation. Since these long range potentials  are often singular at the origin,
it has been argued that the short range interaction should not be represented
by a  3-dimensional delta-function at the origin. Birse et al.
\cite{Birse} choose a delta-shell potential and Beane et al. \cite{Beane}
suggest that a simple attractive square well represents a ``smeared out''
delta-function potential. In either case, it is said, the details of the short
distance regularization should not matter; the low energy aspects of the system
should be invariant in the same way under suitable changes of the short range
potential.

For completeness, we list other regularization and renormalization schemes which do not follow from the separation of scales of an ETF, but also have been applied to the inverse square potential description of physical systems. The problem of a neutral atom interacting with a charged wire \cite{Sch}, relevant to the development of nanoscale atom optical devices, has been treated with the method of self-adjoint extensions \cite{BaCo}. A short distance cutoff scheme which renormalizes the strength of the $1/r^2$ potential, yields a critical dipole moment that has been confronted with the experimental capture of electrons by dipole molecules and formation of anions as an example of quantum mechanical symmetry breaking  \cite{cam},\cite{CoH}. These alternative regularizations are not the subject of our present investigation.

In this note, we follow the regularization method of Beane et al.~\cite{Beane}
to obtain analytically the renormalization group behavior of the coupling
constant of the short range attractive square well they use to regularize the
long range inverse square potential. The $ 1/r^2$ potential is on the boundary
between singular and regular potentials and thus does or does not require a
self-adjoint extension, depending on the strength of the interaction.  More
interesting from the EFT point of view is the Efimov observation \cite{Efim}
that the low energy behavior of three-body systems is determined by a long range
three-body effective interaction of the form $1/R^2$ where $R$ is built from
the relative distances between the particles. Thus, the two-body inverse square
potential is the analogue to the interaction in a three-body system in the
limit of zero energy resonance (and infinite two-body scattering lengths).

A further advantage of our analytic approach to the ``EFT style''
renormalization of the inverse square potential is that we can then more
readily make contact with the mathematically rigorous and well studied approach
to regularization via self-adjoint extensions. The theory of self-adjoint
extensions of Hamiltonians underlies the first discussions of limit cycle
behavior in three-body systems~\cite{AHW}. The self-adjoint extensions of the
inverse square potential are well known \cite{Meetz} and can be compared with
the results of our study, which we now begin. 

 The  starting point of our study is the
$s-$wave reduced radial Schroedinger equation for one particle of mass $m$ in
the external potential $V(r)$: 
\begin{equation}
\left( \frac{d^2}{dr^2} - 2mV(r) + k^2 \right) \psi = 0
\end{equation}
where $V(r)$ is given by \cite{Beane}:
\begin{equation}
V(r) = -\frac{\alpha_s \theta (R-r)}{R^2} - \frac{\alpha \theta(r-R)}{r^2}
\,\,\,\,\,\,
(\alpha_s, \alpha > 0).
\end{equation}
That is, the long range attractive $\alpha/r^2$ second term in eq. (2) is
cutoff at a short distance  radius $R$ by an attractive square well.   As in
\cite{Beane}, we first solve eq. (1) for the zero energy solution  $(k=0)$ $
\psi_o$.  It is given by:
\begin{eqnarray}
\psi_o (r) &=& A \frac{r^{1/2}}{{r_o}^{1/2}}\cos \left( \nu \ln\frac{r}{r_o} +
 \phi_o \right) \hspace{.75in} \,\,\,\,  r>R \\ 
\psi_o (r) &=& A \frac{\cos (\nu \ln\frac{R}{r_o} + \phi_o )}{\sin (K_oR)}
\left( \frac{R^{1/2}}{{r_o}^{1/2}}\right) \sin K_or \,\,\,\,\ \   r<R
\end{eqnarray}
where $ \nu = {(2m\alpha - 1/4)}^{1/2}$, $ \phi_o$ is the zero energy phase
\cite{Beane}, ${K_o}^2 = (2m\alpha_s)/{R^2}$ and $r_o$ is an arbitrary scale.

 The usual matching condition of the wave function and its derivative at
$r=R$ then yields:
\begin{equation} \label{trans}
{(2m\alpha_s)}^{1/2} \cot\{{(2m\alpha_s)}^{1/2}\} = \frac{1}{2} - \nu \tan 
\left( \nu\ln \left( \frac{R}{r_o}\right) + \phi_o \right)
\end{equation}
Following \cite{Beane} we now consider eq. (\ref{trans}) to be a transcendental
equation {\it defining} the value of the short range coupling constant
$\alpha_s$. This equation is of the form $\beta \cot\beta = 1/\omega$ and can
be solved exactly in closed form using a method based upon the solution to the Riemann problem in complex variable analysis \cite{Mus}. 

 The solution to eq.(\ref{trans}) then turns out to be \cite{Burn}:
\begin{eqnarray}
\beta_o &=& \pm \frac{{(\omega - 1)}^{1/2}}{\omega}
\exp\left(\frac{1}{\pi}\int_0^1 \arg \Lambda_o(t)\frac {dt}{t} \right) \, ,
\,\,\,\, \omega >0 \\
\beta_n &=& \pm n\pi \exp \left(\frac{1}{\pi}\int_0^1 \arg 
\Omega_n(t)\frac {dt}{t} \right)
\, , \,\,\,  -\infty<\omega < +\infty,\; n= 1,2,... 
\end{eqnarray}
where:
\begin{eqnarray}
\beta &=& {(2m\alpha_s)}^{1/2}\\
\frac{1}{\omega} &=& \frac{1}{2} - \nu \tan \left(\nu \ln 
\left( \frac{R}{r_o} \right) + \phi_o \right) \\
\Lambda_o(t) &=& \lambda (t) + \textstyle \frac{1}{2}\omega t i \pi \\
\lambda(t) &=& 1+\textstyle \frac{1}{2}\omega t \ln \frac {1-t}{1+t}\\
\Omega_n(t) &=& {\Lambda_o(t)}^2 + n^2{\pi}^2 {\omega}^2t^2 
\end{eqnarray}
Formula (9) does not restrict $\omega$ to be positive, so that all the solutions that we consider follow from the positive branch of (7). If we interpret eq. (\ref{trans}) to be an equation
for the running coupling constant $\beta =  {(2m\alpha_s)}^{1/2}$, this amounts to requiring that $\beta $ must be defined for all values of ${R}/{r_o}$ except for those points  where $\omega$ (or its inverse) vanish. This is potentially important when comparing our analytic solution with {\it numerical} solutions to eq. (5). Indeed, numerical investigations in general will mix solutions that only exist in a limited range of ${R}/{r_o}$-values ($\beta _o$ in eq.(6)) with solutions ($\beta _n$ in eq.(7)) which we consider to be the only physically relevant solutions. From now on, we shall therefore use $\beta $ to mean only a solution of eq. (7), and suppress the subscript $n$ when it is not needed for the discussion. We then find
from eq. (7) that eq. (\ref{trans}) has infinitely many roots, in agreement
with \cite{Beane}. We have plotted $\beta $ in eq. (\ref{trans}) as a function of
$\ln x$  $ (x = {R}/{r_o})$  for fixed $ \phi_o =1.0$, and $n=1$ in  figures 1 and
2.  The strength of the long range potential $\alpha/r^2$  increases with
succeeding figures; $\nu =0.5$ in fig. 1, $\nu = 3.0$ in fig. 2. The running coupling 
$\beta$ exhibits a limit cycle behavior for all values of $\nu > 0$; the period 
becomes smaller as the strength of the attractive $\alpha/r^2$ potential
increases, according to the argument ($ \nu \ln x + \phi_o $) of the tangent function in the source term of eq. (5). The behavior described by $\beta$, for large enough $\nu$, is of a
``sawtooth" type, with a periodic sharp increase of the value of the coupling
constant with decreasing values of  $\ln x$. This ``increase'' is actually a genuine discontinuity of $\beta $ at the zeros of $1/\omega $ and must be a multiple of $\pi $. Indeed, a discontinuity can only occur at a zero of $\cot \beta $ in order that $\beta \cot \beta $ be  continuous at all points where $1/\omega $ is a continuous function. Once $\beta $ has reached the smallest positive zero ($\beta = \pi /2) $, for some $x-$value, it increases by $\pi $ as $x$ is further decreased. Altering the zero energy phase shift
$\phi_o$ from 0 through 2 (and keeping n=1 and $\nu=3$) does not qualitatively
change the appearance of the pattern of Fig. 2: the discontinuity moves to lower $x$ , but the magnitude of the discontinuity remains the same multiple of $\pi$. This feature can be
traced to the periodicity of the right hand side of  eq. (\ref{trans}) with
respect to $\phi_o$. The branches which
correspond to roots with $n > 1$ are  qualitatively similar to Fig 2., but the
slow fall-off with decreasing $x$  seen in Fig. 2 increases and the ``saw-tooth" appearance
becomes more of a rounded off square wave. This feature of the solutions is illustrated by Fig. 3 which plots $\beta $ for the same strength and initial phase $\nu=3$, $ \phi_o = 1$ as Fig. 2, but n has increased to 16. The magnitude of the discontinuity of this $\beta _{16}$ remains $\pi $, however.

 Finally we plot as Fig. 4 the analytical solution of 
eq. (\ref{trans}) for the  values $ \nu = 2.0$, $ \phi_o =0.0$ to compare with
the numerical solutions of eq. (\ref{trans}), with the same input, displayed in
Figure 1 of Ref. \cite{Beane}.  The latter numerical solution is in excellent agreement
with the analytical solution presented here, if one allows  a solution to go from a higher branch ($n=2$) to the next lower branch ($n=1$) as it crosses a zero of $1/\omega$. However, as discussed above and shown on Fig.4, $\beta _1 $ itself must ultimately increase by $\pi $ after reaching its lowest value $\pi/2$, thus exhibiting a limit cycle behavior of the solution.It is evident that the limit cycle behavior of the solutions of equation (5) is a consequence of the requirement that a solution $\beta_n$ be defined for all
     values of $R/r_o$ for which omega or its inverse is nonzero. This requirement {\it includes solutions for $R \rightarrow 0$} and therefore corresponds to our
     understanding of the emulation proposed in ref [3] of the contact term which encapsulates the short range dynamics of an EFT.  From Figures 1-3, it is clear
     that the limit cycle behavior of a given branch continues to the left as $R \rightarrow 0$ and lnx becomes arbitrarily small.  Consider, however, the behavior
     for small $R$ of a numerical solution of ref. [3], shown in Figure 4 which segues smoothly between different branches of the analytic solutions of equation
     (5) as it crosses a zero of $1/\omega$.  For some value of negative ln x, as $R$ becomes arbitrarily small, the numerical solution must pass to the $\beta_0$
     solution.  But, as we have noted, this solution exists for only a limited range of $R/r_o$ values. Thus, the two numerical solutions of Fig. 4 would appear to
     not be defined for arbitrarily small $R/r_o$.  One could, however, choose another numerical solution corresponding to a higher value of $n$ which does not
     pass to $\beta_0$ on the way from large to arbitrarily small $R/r_o$ and avoid this problem. This exercise need not be performed, however, with our
     requirement of a well defined solution for all values of $R/r_o$ for which omega or its inverse is nonzero.

A motivation for both the study of Ref. \cite{Beane} and the present
discussion is the expectation that the two-body inverse square potential is the
analogue to the interaction in a three-body system in the limit of zero energy
resonances.  Indeed, the  approximate  solution of eq. (5) displayed in eq. 8
of \cite{Beane} is quite similar to the equation which describes the running of
the three-body counterterm of the pion-less three-nucleon EFT's of Refs.
\cite{BHK,EFTmisc}. Both equations have poles and the three-body counterterm
seems to reach arbitrarily high values.  We, to the contrary, find no poles in
the analytic solutions of eq. (5). Furthermore, no  evidence of multiple
branches was found in the renormalized pion-less three-nucleon problem. The
renormalization of short distance physics in these two problems needs more
understanding in light of the results of Efimov~\cite{Efim}.

 \begin{figure}
\centering
\includegraphics{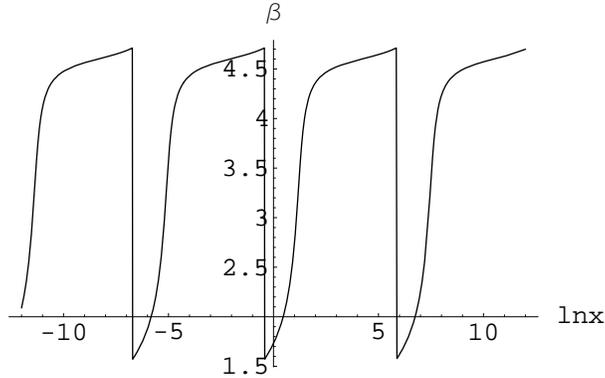}
\caption{The running coupling constant $\beta$ as a function of $\ln x = \ln
\frac{R}{r_o}$ for $ \phi_o =1.0$, $\nu = 0.5$, $n=1$}
% \label{}
 \end{figure}

 \begin{figure}
\centering
\includegraphics{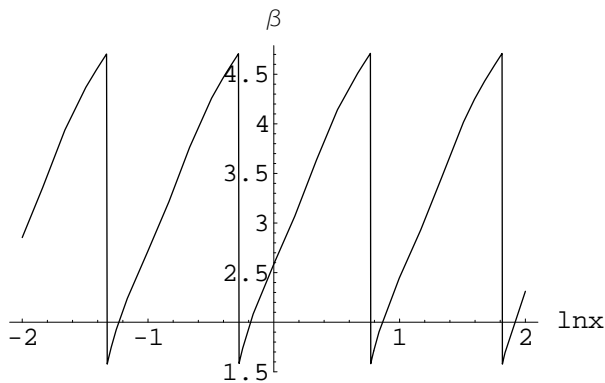}
\caption{The running coupling constant $\beta$ as a function of $\ln x $ for $
\phi_o =1.0$, $\nu = 3.0$, $n=1$}
% \label{}
 \end{figure}

 \begin{figure}
\centering
\includegraphics{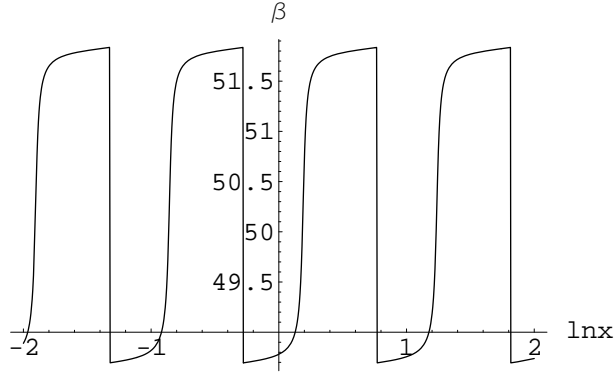}
\caption{ $\beta$ as a function of $\ln x $ for $ \phi_o =1.0$, $\nu = 3.0, n
= 16$} 
% \label{}
 \end{figure}

\begin{figure}
\centering
\includegraphics{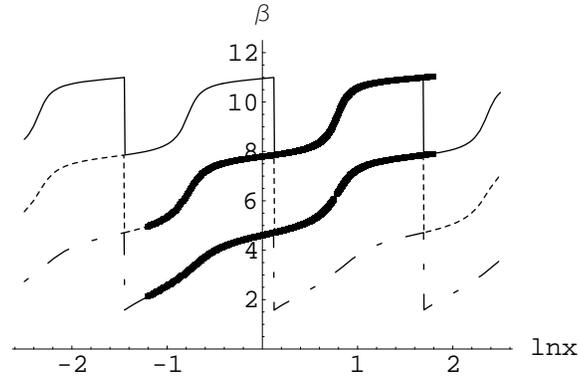}
\caption{ $\beta$ as a function of $\ln x $ for $ \phi_o =0.0$, $\nu = 2.0$.
The three branches, from bottom to top, correspond to n=1 (dash-dotted
curve), n=2 (dotted curve) and n=3 (solid curve).  As explained in the text, the discontinuity in $\beta$ is always $\pi$ for each curve.  The two thicker curves are two
numerical solutions for the same parameters taken from ref. [3], as discussed in the text.}
\label{}
 \end{figure}

Now we turn to the bound state aspects of the related two problems (three-body
system with contact potentials and the $1/r^2$ singular potential) and again
find discrepancies.  First we show that the regularization method that was used
to solve the Schroedinger equation (1) with potential (2) amounts to specifying
a particular self-adjoint extension in Case's solution of the bound state
(B.S.) spectrum of the attractive singular $ {1}/{r^2}$
potential~\cite{Meetz,Case}.  In order to do this, we note that the B.S.
wavefunction that solves eqs. (1)-(2)  is given by:

\begin{eqnarray}
\psi &=& Cr^{1/2}K_{i\nu}(kr)\,\,\,\, r>R\\
\psi &=& C'\sin(Kr)\ \ \ \ \  r<R
\end{eqnarray}
where:
\begin{equation}
K^2 = \frac{2m\alpha_s}{R^2} - k^2
\end{equation} 
and $C$ and $C'$ are constants.

For $kR <<1$, the matching condition now
gives (one still has $KR = {(2m\alpha_s)}^{1/2}$ in that limit):
\begin{equation} \label{case}
k= \frac{1}{2r_o}\exp \frac{\phi_o + \arg \Gamma (1+i\nu) - (n+1/2)\pi }{\nu}
\,\,\,\,\,\,\,\,  n=0, \pm 1, \pm 2, ....
\end{equation}
where we have used eq. (\ref{trans}) together with the small-r behavior
 of $\psi(r)$ \cite{PP}:
\begin{equation}
\psi(r)  \simeq r^{1/2}\sin \left( \nu \log\frac{kr}{2}-\arg\Gamma
(1+i\nu)\right)\, .
\end{equation}

 The spectrum given in eq. (\ref{case}) is essentially the spectrum given by
  Case \cite{Case}, where Case's arbitrary phase (which fixes the self-adjoint
  extension)  $B$ is now given by:

\begin{equation}
B= \phi_o + \arg \Gamma (1+i\nu)  
\end{equation}
It is important to note that the binding energy $E_B = (k^2)/{2m}$, after this
regularization, no longer
depends on the cut-off radius $R$ (for $kR << 1$) but instead on the arbitrary
scale $r_o$. Thus, fixing the zero energy phase of the wavefunction $\phi_o$
removes the cut-off dependence of the B.S. spectrum for $kR << 1$ . Our result
explicitly shows that the physical interpretation of the phase characterizing
the self-adjoint extension of the Hamiltonian indeed can be found from a 
renormalization of the short-range coupling constant in the regularization
method described in Ref. \cite{Beane}. By the same token, it illustrates why a
cut-off method with a constant strength fails \cite{Meetz} to provide a
physical meaning for this arbitrary phase.

Note, however, that the ground state of the $1/r^2$ potential remains at
negative infinity, and no renormalization of the bound state spectrum  has been
achieved. Contrast this result of the EFT style renormalization of
Ref.~\cite{Beane} with the restoration of the lower bound of the pion-less
three-body problem obtained in Refs.~\cite{BHK,AHW}.  A clue to this
discrepancy may lie in the distinction between the contact interaction used in
Ref. 2 and the ``EFT" type \cite{Beane} regularization of the  conventional
3-dimensional delta function. That is, the attractive square well in eq. (2)
does not provide a unique way of regularizing (``smearing out'') a
3-dimensional delta function, and its limit when $ R \rightarrow 0$ is not
 the contact interaction discussed in Ref. 2 and
14. In that respect, it would be quite interesting to reexamine the solution of
the Schroedinger equation with a singular $\alpha/r^2$ potential in conjunction
with local realizations of the contact interaction of Ref. 14 implemented by
Kruppa, Varga and Revai ~\cite{KVR}. Such a study might throw additional light
on the corresponding renormalization group flow properties in the 3-body
problem.

\section*{Acknowledgments}

The work of M.B. was supported by the National Fund for Scientific Research,
Belgium and that of S.A.C. by NSF grant PHY-0070938.  We thank Silas Beane for
providing us with details of the numerical calculations of Ref. ~\cite{Beane}
and Mary Alberg for a communication about our plots of the analytical solutions.

\end{document}